\documentclass[conference]{IEEEtran}
\usepackage{multirow}
\ifCLASSINFOpdf
   \usepackage[pdftex]{graphicx}
   \graphicspath{{../pdf/}{../jpeg/}}
   \DeclareGraphicsExtensions{.pdf,.jpeg,.png}
\else
  \usepackage[dvips]{graphicx}
  \graphicspath{{../eps/}}
  \DeclareGraphicsExtensions{.eps}
\fi
\usepackage{multirow}
\usepackage{booktabs}

\begin{document}
%

\title{Noise Invariant Frame Selection: A Simple Method to Address the Background Noise Problem for Text-independent Speaker Verification}

\author{\IEEEauthorblockN{Siyang Song}
\IEEEauthorblockA{ School of Computer Science\\
University of Nottingham, UK\\
Email: Siyang.Song@nottingham.ac.uk}
\and
\IEEEauthorblockN{Shuimei Zhang}
\IEEEauthorblockA{Department of Electrical and Computer Engineering\\
Temple University, USA\\
Email: tug63690@temple.edu}
\and
\IEEEauthorblockN{Bj\"{o}rn W. Schuller}
\IEEEauthorblockA{ Group on Language, Audio and Music\\
Imperial College London,  UK\\
Embedded Intelligence for Health Care\\
and Wellbeing\\
Augsburg University, Germany\\
Email: schuller@ieee.org}
\and
\IEEEauthorblockN{Linlin Shen}
\IEEEauthorblockA{School of Computer Science and \\
Software Engineering\\
Shenzhen University, China\\
Email: llshen@szu.edu.cn}
\and
\IEEEauthorblockN{Michel Valstar}
\IEEEauthorblockA{School of Computer Science\\
University of Nottingham, UK\\
Email: Michel.Valstar@nottingham.ac.uk}}


%


\maketitle

\begin{abstract}
The performance of speaker-related systems usually degrades heavily in practical applications largely due to the presence of background noise. To improve the robustness of such systems in unknown noisy environments, this paper proposes a simple pre-processing method called Noise Invariant Frame Selection (NIFS). Based on several noisy constraints, it selects noise invariant frames from utterances to represent speakers. Experiments conducted on the TIMIT database showed that the NIFS can significantly improve the performance of Vector Quantization (VQ), Gaussian Mixture Model-Universal Background Model (GMM-UBM) and i-vector-based speaker verification systems in different unknown noisy environments with different SNRs, in comparison to their baselines. Meanwhile, the proposed NIFS-based speaker verification systems achieves similar performance when we change the constraints (hyper-parameters) or features, which indicates that it is robust and easy to reproduce. Since NIFS is designed as a general algorithm, it could be further applied to other similar tasks.
\end{abstract}


%
\IEEEpeerreviewmaketitle

\section{Introduction}

\noindent Speaker verification is the prime example of a speaker-related task, that is, a task for which the data is heavily correlated with speakers. For speaker-related tasks, uncertainty in features can be represented by several speaker models, among which Vector Quantization (VQ), Gaussian mixture models (GMMs) \cite{reynolds1995robust} and i-vector \cite{campbell2006support} are the most successful examples proposed in the past decades. Recently, deep Neural Networks (DNNs), especially Convolution Neural Networks (CNNs)  also have been widely and successfully applied to extract deep features to represent speakers \cite{sercu2016very,matvejka2016analysis,ranjan2017improved}. 

However, in practical applications, speaker verification performance degrades heavily in noisy environments due to the acoustic mismatch between clean training conditions and noisy test conditions \cite{ganapathy2014robust}. Many approaches have been proposed to address this issue: speech enhancement, feature enhancement, model adaptation, etc. Speech enhancement attempts to enhance the signal using the noise information obtained from speech or prior knowledge. To date, enhancement mainly focuses on filtering techniques such as Kalman filtering or spectral subtraction \cite{drygajlo1998speaker}, compensation parallel model combination (PMC) \cite{matsui1996speaker}, Jacobian environmental adaptation \cite{sagayama1997jacobian}, and recently DNNs \cite{mclaren2014application,hruz2017convolutional,watanabe2017student}. Model adaptation is another successful approach, which adjusts parameters of speaker models using different training data while keeping the observation stable \cite{seltzer2013investigation}. For example, hybrid Neural Network/Hidden Markov Model \cite{xue2016speaker}, Linear Spline Interpolation (LSI) \cite{seltzer2010acoustic}, CNNs \cite{ochiai2017cumulative} have been adopted in recent years. Among these adaptation methods, the most successful one is the Universal Background Model \cite{reynolds2000speaker}, which is a universal GMM trained with large amounts of voice data from different speakers. 

All approaches have achieved excellent performance in addressing noise problems under certain conditions, but none of them take the data quality into consideration. In real world application, speakers can be affected by a variety of biological, environmental, social, or cognitive factors (human factors) when they are talking, resulting in distortion of feature vectors extracted from utterances. Although background noise may remain stable over a whole utterance, it may pose different impacts on different frames. For example, some frames may be affected less than others, and would thus be of relatively high quality. In this paper, we propose a pre-processing method called Noise Invariant Frame Selection (NIFS) for selecting a subset of data that is robust to various background noises. Speaker verification is used as the case study here to evaluate the usefulness of the NIFS, but the method could be applied to other tasks that are impacted by noise in the input signals, even they are non-audio signals, i.g. object recognition \cite{Xia20083D} or image retrieval\cite{Song2015A}. The performance of NIFS for speaker verification is evaluated on the TIMIT database \cite{lamel1989speech}.

The rest of this paper is structured as follows: In Section 2, the proposed Noise Invariant Frame Selection algorithm is explained in detail. The experimental setup and results are shown in Section 3. The last section is devoted to the conclusion.

\section{RELATED WORK}

\noindent Previously, a number of works have been investigated in frame selection for acoustic-related tasks. Aiming at selecting task-specified frames, task-specified constraints or criterion are always designed and employed.

Dutoit et al. \cite{dutoit2007towards} adopted the Viterbi algorithm to select a sequence of frames from the target database, which try to minimize a distance between selected frames and the output sequence mapped by their GMM mapping conversion function. The result showed that the combination of mapping and frame selection generate the best results among three experimental systems of their paper.

Ventura et al. \cite{ventura2015audio} presented an audio parameterization method for acoustic recognition of bird species using integrated frame selection method. To be more precise, the proposed frame selection method employed morphological filtering applied on the spectrogram in MFCC algorithm. It allows to exclude from further processing certain audio events, which otherwise could cause misclassification errors. The experiment results for identifying 40 bird species proved its advantages in both accuracy and speed.

By normalizing conventional minimum-redundancy maximum-relevancy (mRMR), Jung et al. \cite{jung2009normalized} proposed the NmRMR criterion. They first extracted features from frames to train an initial feature model. Then feature frames used for the training and test are selected by meeting the NmRMR criterion. This selected frames are expected to have minimum-redundancy within selected feature frames and maximum-relevancy to speaker models. The experiment results verified that the selected frames can enhance the performance of speaker verification system.

Meanwhile, some other researchers have applied more than one constraints to select frames and fusing selection results later.

Bocklet et al. \cite{bocklet2009speaker} proposed a framework to select eight subsets of MFCC feature vectors from the original speech for speaker recognition based on eight different syllable constraints. Then, linear logistic regression is utilized to combine selection results at the score level. The experiment conducted by GMM-UBM system revealed that the proposed frame selection method improved the performance of the baseline system. 

Based on hybrid technique, the framework proposed by Prasad et al. \cite{prasad2017frame} has two branches, where the first one utilizes voice activity detection (VAD) to discard Non-speech frames and conventional Fixed Frame Rate (FFR) to select frames from selected active speech part. The second branch select frames according to the changes in the temporal characteristics of speech based on Variable Frame Rate (VFR) analysis. Finally, the selected frames from both branches are concatenated for the further processing.

Nematollahi et al. \cite{nematollahi2016speaker} extracted linear predictive coefficients (LPC), Gain and LP residual from each frame and then proposed three different ways to weight them. After that, the sum of the weighted scores is utilized for frame selection. Since the higher weight stands for the better speaker discrimination ability, those frames with lower weight are discarded.

\begin{figure*}[!t]
	\centering
	\includegraphics[scale=0.9]{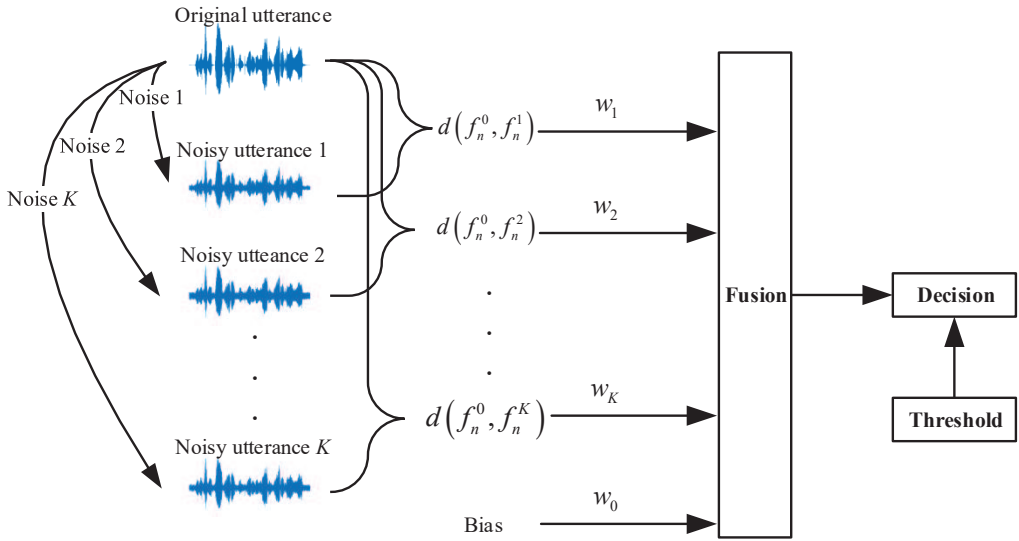}
	\caption{The Principle of NIFS. The weighting part and fusion part are corresponding to Equation (2) and Equation (3) respectively.}
	\label{NIFS}
\end{figure*}

\section{METHODOLOGY}

\noindent Besides the model architecture, training data is also crucial for a model's generalization ability. A subset of training data may train a model that has a better performance than the original set. The key question is how to select the optimized subset. Motivated by this, we propose a NIFS framework that aims to select noise invariant frames from utterances.

\subsection{Robust frame selection}

\noindent According to Fig.\ref{NIFS}, $K$ kinds of additive noise are taken into account as the constraints. By adding each additive noise to the original utterance respectively, $K$ noisy utterances with the same number of frames are generated. It should be noted that a original utterance is not equal to the clean utterance. Then, $K+1$ sets of feature vectors can be extracted from these noisy utterances and the original utterance. For each frame of the original utterance, the distances between its feature vector and feature vectors extracted from the corresponding $K$ noisy utterances are calculated separately, which can be denoted as
$$
D \left( n \right) =\left\{ d\left(f_n^o,f_n^1 \right), d\left(f_n^o,f_n^2 \right) \dots d\left(f_n^o,f_n^K \right) \right\} \eqno{(1)}
$$
where $d\left(f_n^o,f_n^k \right)$ is the distance between the feature vector extracted from the $n_{th}$ frame of the original utterance and the feature vector extracted from its corresponding frame of the $k_{th}$ noisy utterance, where $k = 1,2, \dots, K$. Here, the Euclidean distance is employed as the measure of distortion but other measures can be adopted for different tasks. Since more than one noise constraint is utilized and their impacts on the same frame may different, the weights and bias are applied to represent this uncertainty. As a result, the score can be calculated from the distance between an original frame $f_n^o$ and a noisy frame $f_n^k$ by
$$
S\left( f_n^o, f_n^k\right) = w_k \times d\left(f_n^o,f_n^k \right) \eqno{(2)}
$$
where $w_k$ denotes the weight of the $k$th noisy utterance. After scores between those clean frames and their corresponding noisy frames are calculated, the final score of $f_n$ can be obtained by fusing the scores of all constraints, which can be factorized as
$$
S_{sum}\left(f_n \right)=\sum_{k=1}^K S\left( f_n^o, f_n^k\right)  + w_0  \eqno{(3)}
$$
where $w_0$ denotes the bias.

Consequently, frames that are less distorted by noise will have lower scores. By ranking all original frames in ascending order based on their final scores (Equation (4)), a new set of frames that are less sensitive to noise can be generated by selecting top ranked frames.
$$
SF= \left\{f_n \mid  S_{sum}\left(f_n \right) \in TOP(w)     \right\} \eqno{(4)}
$$

Unfortunately, the weights and bias of Equation (2), (3) and (4) would take quite a long time to be optimized. Therefore, a simple function for fusing distances is proposed in Equation (5). It ranks frames in ascending order for each constraint individually and the final subset can be obtained by selecting the intersection of them.
$$
SF= RS_1 \left(w \right)\bigcap RS_2 \left(w \right)\bigcap \dots \bigcap  RS_K\left(w \right) \eqno{(5)}
$$ 
where $RS_k$ is a ranked subset obtained by $k_{th}$ constraint, $w$ is the percentage of frames that $RS_i$ kept and $SF$ is the final selected frame set. This formulation can reduce the number of parameters from $K+1$ ($w_1, w_2,  \dots, w_k$ and $w_0$) to only one ($w$) because all weights are set as the same value while the bias is set to zero. The reason of using $w$ rather than a fixed threshold is that by using it, the NIFS framework can provide task-specified frame sets by controlling the robust degree and the number of the selected frames.

Given that there are $m$ utterances for training each speaker and $K$ noisy constraints are utilized ($m > 1$, $K > 1$), the time complexity of parameter optimization for both methods are analyzed as follows. For Equation (4), a linear search is applied to find the best threshold $w$, while we make use of Batch Gradient Descent to obtain the optimized weights and bias in Equation (2) and (3) for each threshold $w$. Since the time complexity of linear search and Batch Gradient Descent are $O(n)$ and $O(mK)$, the complexity of this method is $O(nmK)$. Fortunately, for the simple version (Equation (5)), only linear search needs to be adopted for searching the best threshold $w$, for which time complexity is $O(n)$.

\begin{figure*}[!t]
	\centering
	\includegraphics[scale=0.55]{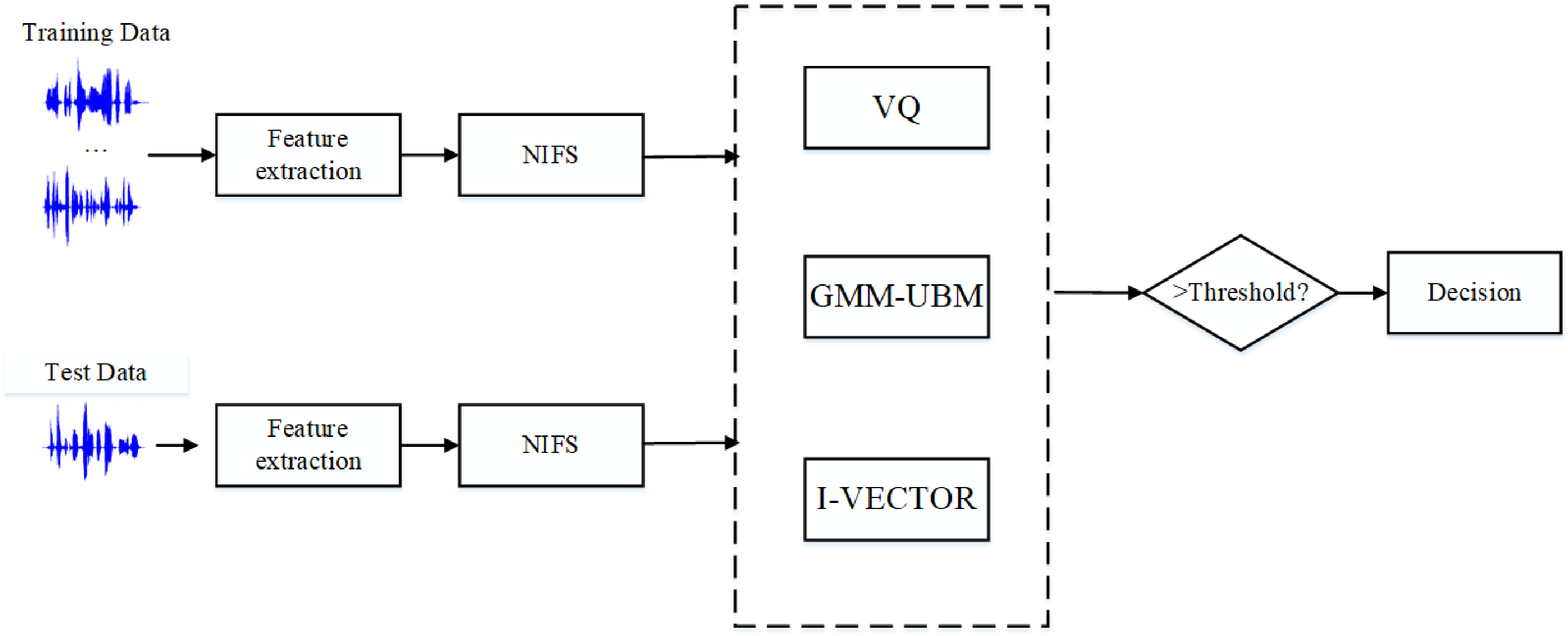}
	\caption{The framework of speaker verification experiments}
	\label{system}
\end{figure*}

\subsection{Speaker verification using robust frames}

The NIFS algorithm is easy to be applied to any single utterance for selecting noise invariant frames, and thus we utilize speaker verification to validate the effectiveness of it.

In terms of applying the NIFS to speaker verification as a front-end, it can be adopted in both training and test phases. When applying the NIFS to training phase, it selects those robust training frames which are then used to train speaker models directly. When applying the NIFS to both training and test phase, it first selects robust training frames to train speaker models. Then, when a test utterance appeared, NIFS also selects robust frames from it and the input these frames to speaker models for testing.

\section{EXPERIMENTAL RESULTS}

\begin{table*}[h]
	\caption{THE EER OF BASELINES AND THEIR NIFS SYSTEMS (\%)}
	\label{table_1}
	\begin{center}
		\scalebox{1}[1]{
			\begin{tabular}{ c c c c c c c c}
				\toprule
				\multirow{2}{*}{Test conditions} & \multirow{2}{*}{SNR(dB)} &
				\multicolumn{6}{c}{Speaker Verification system}\\
				\cmidrule{3-8}
				&  & VQ baseline & NIFS-VQ & GMM-UBM baseline & NIFS-GMM-UBM & I-vector baseline & NIFS-I-vector \\
				\midrule
				clean &            & 16.9 & 8.3  & 5.4  & 5.6  & 6.2  &  5.0\\
				\hline
				& 25  & 23.1 & 12.5 & 6.9  & 6.6  & 9.6   & 8.5\\
				
				Factory      & 20  & 26.9 & 15.6 & 9.4  & 8.1  & 10.6  & 10.5 \\
				
				& 15  & 32.8 & 22.5 & 10.8 & 10.6 & 13.8  & 11.8\\
				\hline
				& 25  & 18.1 & 11.6 & 6.6  & 6.2  & 7.8   & 5.6\\
				
				Machinegun   & 20  & 20.4 & 12.9 & 6.5  & 6.3  & 8.8   & 7.5 \\
				
				& 15  & 24.5 & 14.7 & 6.9  & 7.5  & 9.3    & 8.5\\
				\hline
				& 25  & 18.2 & 11.9 & 8.8  & 8.5  & 8     & 7.9\\
				
				Volvo        & 20 & 20.6 & 15.0  & 8.8  & 7.7  & 8.8   & 8.1 \\
				
				& 15 & 26.5 & 18.8  & 8.4  & 7.9  & 10.4  & 10.3\\
				\hline
				& 25 & 23.1 & 13.1  & 6.3  & 5.8  & 8.8  & 9.3\\
				
				Leopard      & 20 & 26.9 & 14.4  & 7.5  & 6.6  & 11.4  & 8.8 \\
				
				& 15 & 32.7 & 18.3  & 9.1  & 7.1  & 13.5  & 11.2\\
				\hline
				Average &         & 23.9 & 14.6  & 7.8  & 7.3  & 9.8  & 8.7\\
				\bottomrule
			\end{tabular}
		}
	\end{center}
\end{table*}

\subsection{Experimental setup}

\noindent In this paper, a 24-dimensional MFCC feature \cite{davis1980comparison} consisting of 12 MFCC and 12 $\Delta$ MFCC is utilized. Each frame of the utterances is processed by a 25 ms length Hamming window and shifted by 10 ms. The 0th cepstral coefficient is replaced with the log energy. All experiments were carried out on the core condition of the TIMIT database, which contains a total of 6300 voice samples of 630 speakers (438 males and 192 females and each speaker contribute 10 utterances) from 8 major dialect regions in the United States. In our experiments, 80 speakers made up of 57 male speakers and 23 female speakers, were randomly selected (The index of these speakers and the MATLAB code of NIFS can be found in https://github.com/shuimove1234/Noise-Invariant-Frame-Selection).

In the training phase, eight utterances were utilized to train models for each speaker while another two remaining utterances were used for testing (160 test utterances in total). To test the performance of the NIFS in unknown noisy environments, test utterances were corrupted by four types of additive noise including factory noise, leopard noise, machinegun noise and volvo noise, resulting in noisy test utterances at 15, 20, 25 dB SNR. Meanwhile, a clean condition is also introduced, where original test utterances without adding any additive noises were applied. It should be noted that the additive noise used in the test phase were different from the noise constraints used for training. All noise used in this paper is provided by the NOISEX-92 database \cite{varga1993assessment}.  

In terms of speaker models, the codebooks of VQ systems were constructed with 128 clusters while the GMM models had 128 Gaussian mixture components. The universal background model was trained by 70 males and 30 females who were randomly selected in the remaining TIMIT (1000 speech recordings in total). Because the purpose of the experiments is to justify whether the selected frames can better represent the speaker than the original frames, no other pre-processing or post-processing method was applied.

\begin{table*}[h]
	\caption{Equal Error Rate OBTAINED FROM NIFS-IVECTOR SYSTEMS WITH DIFFERENT NOISE COMBINATIONS (\%)}
	\label{table_2}
	\begin{center}
      \scalebox{1}[1]{
		\begin{tabular}{c c c c c c c c c c c c c c}
			\toprule
			\multirow{2}{*}{Test conditions} & \multirow{2}{*}{Baseline}&
			\multicolumn{3}{c}{Noise combination}&&
			\multicolumn{3}{c}{Noise SNR}&&
			\multicolumn{3}{c}{Noise number}\\
			\cmidrule{3-5} \cmidrule{7-9} \cmidrule{11-13}
			&  & 1st & 2nd & 3rd & & 15 & 20 & 25 & & 1 & 2 & 3 \\
			\midrule
			Factory    & 10.6 &10.5 &11.2 &8.2   & &10.6 &10.5 &8.2 & &10.9 &8.8 &10.5 \\
			Machinegun & 8.8  &7.5  &7.2  &7.5   & &7.5  &7.5  &8.1 & &9.3  &7.1 &7.5\\
			Volvo      & 8.8  &8.1  &8.7  &10.6  & &9.3  &8.1  &8.7 & &8.7  &8.1 &8.1\\
			Leopard    & 11.4 &8.8  &9.2  &8.9   & &9.3  &8.8  &8.1 & &11.4 &11.0 &8.8\\
			Average    & 9.9  &8.7  &8.8  &8.8   & &9.2  &8.7  &8.3 & &10.1 &8.8  &8.7\\
			\bottomrule
		\end{tabular}
        }
	\end{center}
\end{table*}

\subsection{{Speaker verification results}}

In this section, three types of noises, including babble noise, white noise and pink noise from the NOISEX-92 database were introduced as the constraints in the training phase of the NIFS, where the SNRs were all set to be 20 dB.

The performances of the NIFS were evaluated using the equal error rate (EER) on three sub-experiments: 1. NIFS and VQ (NIFS-VQ)-based speaker verification; 2. NIFS and GMM-UBM (NIFS-GMM-UBM)-based speaker verification; 3. NIFS and i-vector (NIFS-i-vector)-based speaker verification (see Fig. \ref{system}). To be more specific, the original utterances were processed by the NIFS to yield corresponding selected frame sets, which then are fed to speaker models for training and testing.

\begin{figure*}[!t]
	\centering
	\includegraphics[scale=0.4]{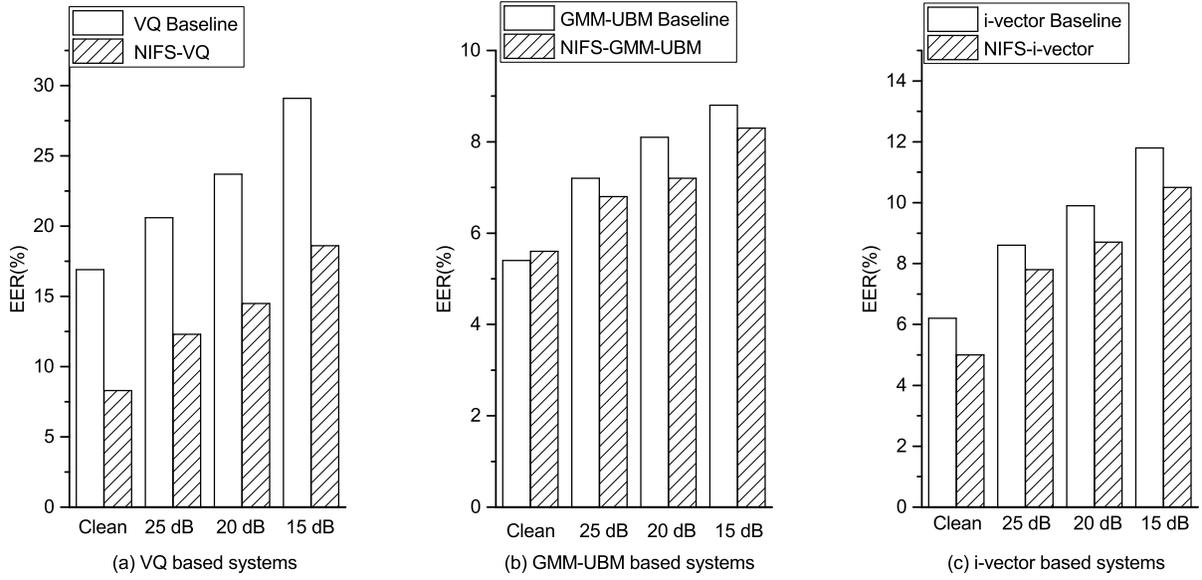}
	\caption{Equal Error Rate comparison between models with and without NIFS.}
	\label{SystemEER}
\end{figure*}

The speaker verification results of these three NIFS-based systems and their corresponding baselines (VQ, GMM-UBM, i-vector) under clean condition and different noisy conditions are presented in Table \ref{table_1}. It is clear that the performance of NIFS-VQ and NIFS-i-vector outperformed their corresponding baselines under all conditions, while NIFS-GMM-UBM only generated worse result in the environment with machinegun noise at 15 dB SNR. This may due to that NIFS decreased the number of training frames, resulting in speaker models were not well-trained. The average relative improvements of NIFS for VQ, GMM-UBM and i-vector baselines in 12 noisy environments are 38.4\%, 7.5\% and 21.7\%, respectively. The result comparison shown in Fig. \ref{SystemEER} illustrated the effectiveness of the proposed NIFS.

\subsection{{Constraints analysis}}

\noindent The constraints used in the NIFS have three main hyper-parameters: the types of noise, the number of noise and the noise SNR. It is interesting to explore the impact of these hyper-parameters on speaker  verification. Motivated by this, the following three experiments were conducted, where i-vector was employed as the speaker model.

\subsubsection{Noise types}

\noindent The purpose of the first experiment is to discover the influence of different types of noise constraints adopted in the NIFS. Besides the combination of three noise types utilized in the aforementioned experiment, we also introduced another two combinations of noise constraints. One is the combination of `buccaneer1', `f16' and `m109' (noise combination 2) and the other is the combination is the `destroyerengine', `destroyerops' and `hfchannel' (noise combination 3). Then, new training data selected by NIFS with noise combination 2 and combination 3 were applied to train another two i-vector models respectively. Finally, well-trained models were tested on four different noisy environments with noise SNR of 20 dB. According to the table \ref{table_2}, although frame sets selected by different noise combination yield different results on four noisy test environments, the average results are similar and all of them are better than the baseline.

\subsubsection{Noise SNR}

\noindent With the same training and testing setup, the second experiment applied the noise combination one with three different SNRs: 15 dB, 20 dB, 25 dB, as the constraints. The result displayed in Table \ref{table_2} demonstrates that the SNR of constraints has certain impact on the quality of the selected frames. Fortunately, all selected frame sets outperformed the original frame set.

\subsubsection{Noise number}

\noindent The objective of the third experiment is to explore the influence of the number of noise constraints used in the NIFS, where the first model only made use of one noise (white noise with SNR of 20 dB) in the selection phase while the second and the third model utilized two (babble and white noises with SNR of 20 dB) and three noises (babble, white and pink noises with SNR of 20 dB) separately. It is clear from the last part of Table \ref{table_2} that using only one noise constraints may not select high quality noise invariant frames. However, when more noise constraints were applied, the frames selected by NIFS can further enhance the performance of i-vector for speaker verification.

\subsection{Feature analysis}

\noindent Since the NIFS is proposed as a general framework, it should be suitable for different features. Therefore, besides the 24 dimensional MFCCs, the last experiment also employed another two features: 39 dimensional MFCC (13MFCC + 13$\Delta$MFCC + 13$\Delta$$\Delta$MFCC) and 60 dimensional MFCC (20MFCC + 20$\Delta$MFCC + 20$\Delta$$\Delta$MFCC). The hyper-parameters for this experiment were set as the same as the experiment conducted in section B. Consequently, another two i-vector models were trained by 39 dimensional MFCCs and 60 dimensional MFCCs. The result displayed in Table \ref{table_3} proved that the NIFS has improved the EER results for all models in the clean and noisy environments, which justified that the proposed NIFS method is a useful front-end that can enhance the speaker verification performance for different features.

The next experiment was conducted to evaluate the usefulness of the NIFS in training and test phase individually. According to Table \ref{table_6}, the systems which either applied NIFS to the test phase or the training phase outperformed the baseline which haven't used it. When applied NIFS to both phases, the system yield better result than either applied it to training phase or test phase separately.

\begin{table}[h]
	\caption{THE EER OBTAINED FROM NIFS-IVECTOR SYSTEMS WITH DIFFERENT DIMENSION OF MFCC (\%)}
	\label{table_3}
	\begin{center}
		\begin{tabular}{c c c c c c c }
			\toprule
			\multirow{3}*{Test conditions} &\multicolumn{6}{c}{MFCC dimension}\\
			\cmidrule{2-7}
			&\multicolumn{2}{c}{24}&\multicolumn{2}{c}{39}&\multicolumn{2}{c}{60}\\
			&w/o &w &w/o &w &w/o &w\\
			\midrule
			Factory    & 10.6 &10.5 & 11.3  & 8.8  &6.9 &6.3 \\
			Machinegun & 8.8  &7.5  & 8.1   & 7.8  &6.3 &4.4 \\
			Volvo      & 8.8  &8.1  & 8.8   & 8.8  &4.5 &4.1\\
			Leopard    & 11.4 &8.8  & 16.9  & 11.7 &9.4 &7.5\\
			Average    & 9.9  &8.7  & 11.3  & 9.3  &6.7 &5.6\\
			\bottomrule
		\end{tabular}
	\end{center}
\end{table}

	\begin{table}[h]
		\caption{THE EER OBTAINED FROM I-VECTOR BASED SYSTEMS WITH APPLYING NIFS IN DIFFERENT PHASES (\%)}
		\label{table_6}
		\begin{center}
			\begin{tabular}{c c c c c }
				\toprule
				\multirow{2}{*}{Noise type} & \multirow{2}{*}{Baseline} & 
				\multicolumn{3}{c}{Phase}\\
				\cmidrule{3-5}	
				&  & Both & Train & Test \\
				\midrule
				Factory    & 10.6  & 10.5  & 9.4   & 10.6 \\
				Machinegun & 8.8   & 7.5   & 7.5   & 6.9 \\
				Volvo      & 8.8   & 8.1   & 8.5   & 7.5 \\
				Leopard    & 11.4  & 8.8   & 10.6  & 10.6\\
				Average    & 9.9   & 8.7   & 9.4   & 8.9\\
				\bottomrule
			\end{tabular}
		\end{center}
	\end{table}

\subsection{{Parameter Analysis}}

\begin{figure}[!t]
	\centering
	\includegraphics[scale=0.35 ]{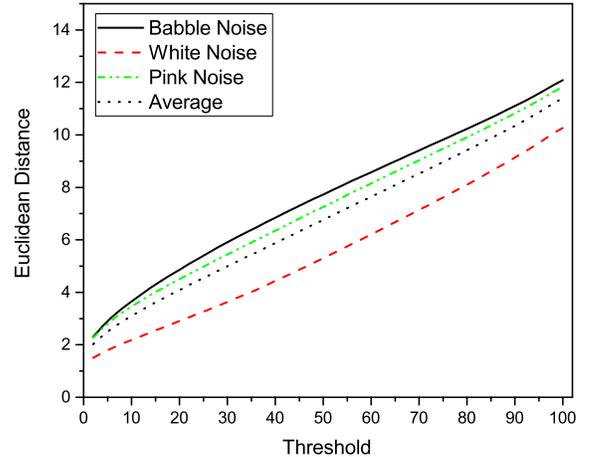}
	\caption{The relation between selection threshold and Euclidean distance }
	\label{Distance}
\end{figure}

\noindent There is clearly a trade-off between the quality of frames and the number of the selected frames. To study this trade-off, the relationship between the threshold $w$ adopted in the Equation (5) and the average Euclidean distance between the selected frames in original utterances and its corresponding noisy utterances is displayed in Fig. \ref{Distance}. The average distance decreased almost linearly with the threshold reducing, which justified that NIFS is able to remove those frames that are easily to be distorted by noises.

\begin{figure*}[!t]
	\centering
	\includegraphics[scale=0.6 ]{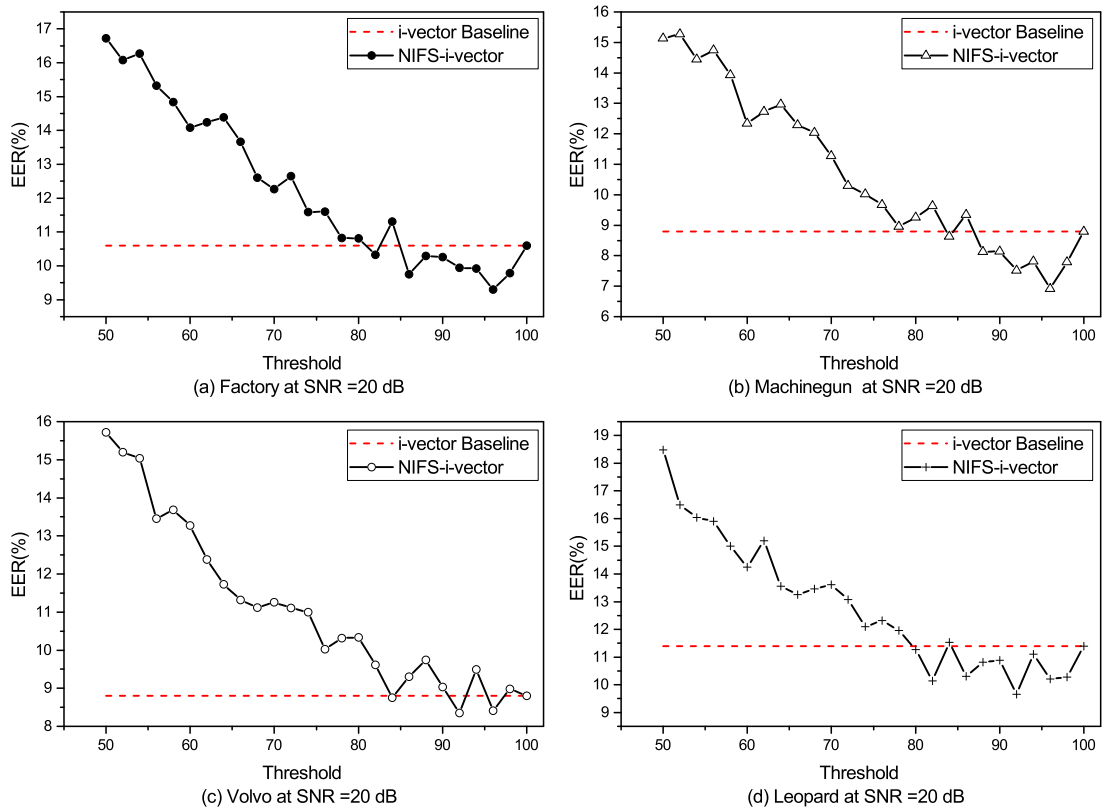}
	\caption{The relation between selection threshold and speaker verification performance}
	\label{eervsepsilon}
\end{figure*}

However, selecting a subset of highly robust frames from the original set will lead to fewer frames being utilized in the training phase. This may negatively affect the performance of models. Hence, we also studied the relationship between the EER result and the threshold $w$ adopted in the Equation (5). According to Fig. \ref{eervsepsilon}, removing those frames that are easiest to be distorted can enhance the EER. However, after a certain percentage ($90 \% - 92 \%$), the benefit of NIFS cannot compensate the negative impact brought by the reduced number of the training data. Fortunately, since there is only one threshold to be determined for the NIFS, it is easy to be optimized.

\section{{CONCLUSION}}

\noindent Background noise poses unequal impacts on different frames of an utterance, where a subset of robust frames may train a better model for acoustic-related tasks. To justify this hypothesis, we have proposed a simple Noise Invariant Frame Selection (NIFS) method of low computational complexity as the front-end for speaker verification. NIFS applies several additive noisy copies of the input utterances as the constraints to select robust frames from utterances to represent speakers. The results show that speaker verification performance is improved under almost all conditions (clean and noisy) for each model by using the proposed NIFS as the front-end. Experiments demonstrated that although changing hyper-parameters of noise constraints of NIFS can affect the quality of the selected frames and result in different speaker verification results, performance still improves under almost all unknown noisy conditions. Experiments also proved that NIFS is useful for both training and testing and it is suitable for different features.

In conclusion, there exist some frames that are relatively robust to different kinds of additive noises and correctly selecting these frames is absolutely essential for good performance. This paper proved that frames selected by the proposed NIFS can train better speaker models than the original training set. Also, it is easy to be reproduced. Although NIFS only has been applied to a speaker verification task in this paper, we believe that it can be easily extended as a pre-processing method to other pattern recognition tasks.

\bibliographystyle{IEEEbib}
\bibliography{refs}

\begin{thebibliography}{10}

\bibitem{reynolds1995robust}
D.~A. Reynolds and R.~C. Rose,
\newblock ``Robust text-independent speaker identification using gaussian
  mixture speaker models,''
\newblock {\em IEEE Trans. Audio, Speech, Lang. Process.}, vol. 3, no. 1, pp.
  72--83, Jan 1995.

\bibitem{campbell2006support}
W.~M. Campbell, D.~E. Sturim, and D.~A. Reynolds,
\newblock ``Support vector machines using gmm supervectors for speaker
  verification,''
\newblock {\em IEEE Signal Process. Lett.}, vol. 13, no. 5, pp. 308--311, May
  2006.

\bibitem{sercu2016very}
T.~Sercu, C.~Puhrsch, B.~Kingsbury, and Y.~LeCun,
\newblock ``Very deep multilingual convolutional neural networks for lvcsr,''
\newblock in {\em Proc. IEEE Int. Conf. Acoust. Speech Signal Process.
  (ICASSP)}, March 2016, pp. 4955--4959.

\bibitem{matvejka2016analysis}
P.~Matějka, O.~Glembek, O.~Novotný, O.~Plchot, F.~Grézl, L.~Burget, and
  J.~H. Cernocký,
\newblock ``Analysis of dnn approaches to speaker identification,''
\newblock in {\em Proc. IEEE Int. Conf. Acoust. Speech Signal Process.
  (ICASSP)}, March 2016, pp. 5100--5104.

\bibitem{ranjan2017improved}
Shivesh Ranjan and John H.~L. Hansen,
\newblock ``Improved gender independent speaker recognition using convolutional
  neural network based bottleneck features,''
\newblock {\em Proc. Interspeech 2017}, pp. 1009--1013, 2017.

\bibitem{ganapathy2014robust}
S.~Ganapathy, S.~H. Mallidi, and H.~Hermansky,
\newblock ``Robust feature extraction using modulation filtering of
  autoregressive models,''
\newblock {\em IEEE/ACM Trans. Audio, Speech, Lang. Process.,}, vol. 22, no. 8,
  pp. 1285--1295, Aug 2014.

\bibitem{drygajlo1998speaker}
A.~Drygajlo and M.~El-Maliki,
\newblock ``Speaker verification in noisy environments with combined spectral
  subtraction and missing feature theory,''
\newblock in {\em Proc. IEEE Int. Conf. Acoust. Speech Signal Process.
  (ICASSP)}, May 1998, vol.~1, pp. 121--124 vol.1.

\bibitem{matsui1996speaker}
T.~Matsui, T.~Kanno, and S.~Furui,
\newblock ``Speaker recognition using hmm composition in noisy environments,''
\newblock {\em Comput. Speech Lang.}, vol. 10, no. 2, pp. 107--116, 1996.

\bibitem{sagayama1997jacobian}
S.~Sagayama, Y.~Yamaguchi, S.~Takahashi, and J.~Takahashi,
\newblock ``Jacobian approach to fast acoustic model adaptation,''
\newblock in {\em Proc. IEEE Int. Conf. Acoust. Speech Signal Process.
  (ICASSP)}, Apr 1997, vol.~2, pp. 835--838 vol.2.

\bibitem{mclaren2014application}
M.~McLaren, Y.~Lei, N.~Scheffer, and L.~Ferrer,
\newblock ``Application of convolutional neural networks to speaker recognition
  in noisy conditions.,''
\newblock in {\em Proc Interspeech}, 2014, pp. 686--690.

\bibitem{hruz2017convolutional}
M.~Hrúz and Z.~Zajíc,
\newblock ``Convolutional neural network for speaker change detection in
  telephone speaker diarization system,''
\newblock in {\em Proc. IEEE Int. Conf. Acoust. Speech Signal Process.
  (ICASSP)}, March 2017, pp. 4945--4949.

\bibitem{watanabe2017student}
S.~Watanabe, T.~Hori, J.~Le Roux, and J.~R. Hershey,
\newblock ``Student-teacher network learning with enhanced features,''
\newblock in {\em 2017 IEEE International Conference on Acoustics, Speech and
  Signal Processing (ICASSP)}, March 2017, pp. 5275--5279.

\bibitem{seltzer2013investigation}
M.~L. Seltzer, D.~Yu, and Y.~Wang,
\newblock ``An investigation of deep neural networks for noise robust speech
  recognition,''
\newblock in {\em Proc. IEEE Int. Conf. Acoust. Speech Signal Process.
  (ICASSP)}, May 2013, pp. 7398--7402.

\bibitem{xue2016speaker}
S.~Xue, H.~Jiang, L.~Dai, and Q.~Liu,
\newblock ``Speaker adaptation of hybrid nn/hmm model for speech recognition
  based on singular value decomposition,''
\newblock {\em Journal of Signal Processing Systems}, vol. 82, no. 2, pp.
  175--185, 2016.

\bibitem{seltzer2010acoustic}
M.~L. Seltzer, A.~Acero, and K.~Kalgaonkar,
\newblock ``Acoustic model adaptation via linear spline interpolation for
  robust speech recognition,''
\newblock in {\em Proc. IEEE Int. Conf. Acoust. Speech Signal Process.
  (ICASSP)}, March 2010, pp. 4550--4553.

\bibitem{ochiai2017cumulative}
T.~Ochiai, M.~Delcroix, K.~Kinoshita, A.~Ogawa, T.~Asami, S.~Katagiri, and
  T.~Nakatani,
\newblock ``Cumulative moving averaged bottleneck speaker vectors for online
  speaker adaptation of cnn-based acoustic models,''
\newblock in {\em Proc. IEEE Int. Conf. Acoust. Speech Signal Process.
  (ICASSP)}, March 2017, pp. 5175--5179.

\bibitem{reynolds2000speaker}
D.~Reynolds, T.~F. Quatieri, and R.~B. Dunn,
\newblock ``Speaker verification using adapted gaussian mixture models,''
\newblock {\em Digit. Signal Process.}, vol. 10, no. 1-3, pp. 19--41, 2000.

\bibitem{Xia20083D}
Shengping Xia and Edwin~R. Hancock,
\newblock ``3d object recognition using hyper-graphs and ranked local invariant
  features,''
\newblock in {\em Joint Iapr International Workshop on Structural, Syntactic,
  and Statistical Pattern Recognition}, 2008, pp. 117--126.

\bibitem{Song2015A}
Siyang Song, Shengping Xia, Zhaosheng Teng, and Shuimei Zhang,
\newblock ``A precise and real-time loop-closure detection for slam using the
  rsom tree,''
\newblock {\em International Journal of Advanced Robotic Systems}, vol. 12, no.
  6, 2015.

\bibitem{lamel1989speech}
L.~F. Lamel, R.~H. Kassel, and S.~Seneff,
\newblock ``Speech database development: Design and analysis of the
  acoustic-phonetic corpus,''
\newblock in {\em Speech Input/Output Assessment and Speech Databases}, 1989.

\bibitem{dutoit2007towards}
T.~Dutoit, A.~Holzapfel, M.~Jottrand, A.~Moinet, J.~Prez, and Y.~Stylianou,
\newblock ``Towards a voice conversion system based on frame selection,''
\newblock in {\em Proc. IEEE Int. Conf. Acoust. Speech Signal Process.
  (ICASSP)}, April 2007, vol.~4, pp. IV--513--IV--516.

\bibitem{ventura2015audio}
Thiago~M Ventura, Allan~G de~Oliveira, Todor~D Ganchev, Josiel~M de~Figueiredo,
  Olaf Jahn, Marinez~I Marques, and Karl-L Schuchmann,
\newblock ``Audio parameterization with robust frame selection for improved
  bird identification,''
\newblock {\em Expert Systems with Applications}, vol. 42, no. 22, pp.
  8463--8471, 2015.

\bibitem{jung2009normalized}
Chi-Sang Jung, Moo-Young Kim, and Hong-Goo Kang,
\newblock ``Normalized minimum-redundancy and maximum-relevancy based feature
  selection for speaker verification systems,''
\newblock in {\em Acoustics, Speech and Signal Processing, 2009. ICASSP 2009.
  IEEE International Conference on}. IEEE, 2009, pp. 4549--4552.

\bibitem{bocklet2009speaker}
T.~Bocklet and E.~Shriberg,
\newblock ``Speaker recognition using syllable-based constraints for cepstral
  frame selection,''
\newblock in {\em Proc. IEEE Int. Conf. Acoust. Speech Signal Process.
  (ICASSP)}, April 2009, pp. 4525--4528.

\bibitem{prasad2017frame}
Swati Prasad, Zheng-Hua Tan, and Ramjee Prasad,
\newblock ``Frame selection for robust speaker identification: A hybrid
  approach,''
\newblock {\em Wireless Personal Communications}, pp. 1--18, 2017.

\bibitem{nematollahi2016speaker}
Mohammad~Ali Nematollahi, SAR Al-Haddad, Shyamala Doraisamy, and Hamurabi
  Gamboa-Rosales,
\newblock ``Speaker frame selection for digital speech watermarking,''
\newblock {\em National Academy Science Letters}, vol. 39, no. 3, pp. 197--201,
  2016.

\bibitem{davis1980comparison}
S.~Davis and P.~Mermelstein,
\newblock ``Comparison of parametric representations for monosyllabic word
  recognition in continuously spoken sentences,''
\newblock {\em IEEE Trans. Acoustics, Speech, Signal Process.}, vol. 28, no. 4,
  pp. 357--366, Aug 1980.

\bibitem{varga1993assessment}
A.~Varga and H.~J.~M. Steeneken,
\newblock ``Assessment for automatic speech recognition: Ii. noisex-92: A
  database and an experiment to study the effect of additive noise on speech
  recognition systems,''
\newblock {\em Speech Commun.}, vol. 12, no. 3, pp. 247--251, 1993.

\end{thebibliography}

\end{document}